\documentclass[%
preprint,
 amsmath,amssymb,
 aps,
]{revtex4-1}

\usepackage{graphicx}
\usepackage{dcolumn}
\usepackage{bm}

\begin{document}

\title{Frequency modulated self-oscillation and phase inertia in a synchronized nanowire mechanical resonator}


\author{T. Barois}
\author{S. Perisanu}
\author{P. Vincent}
\author{S. T. Purcell}
\author{A. Ayari}%
 \email{anthony.ayari@univ-lyon1.fr}
\affiliation{%
Institut Lumi\`ere Mati\`ere, UMR5306 Universit\'e Lyon 1-CNRS, Universit\'e de Lyon 69622 Villeurbanne cedex, France.}%

\date{\today}

\begin{abstract}
Synchronization has been reported for a wide range of self-oscillating systems, but though predicted theoretically for several decades, the experimental realization of phase self-oscillation sometimes called phase trapping in the high driving regime has been studied only recently. We have explored in details the phase dynamics in a synchronized field emission SiC nano-electro-mechanical system with intrinsic feedback. A richer variety of phase behavior has been unambiguously identified implying phase modulation and inertia. This synchronization regime is expected to have implications for the comprehension of the dynamics of interacting self-oscillating networks and for the generation of frequency modulated signals at the nanoscale.
\end{abstract}

\maketitle

A nanoelectromechanical system (NEMS) can be defined as a mobile electrical device at the nanoscale\cite{Craighead} capable for instance of ultrasensitive mass detection\cite{chaste2012}. It has been shown that non-linearities\cite{lifshitz2008nonlinear} can play a major role in such devices\cite{cross2008,PhysRevB.81.165440,eichler_nonlinear_damping}. Self-oscillating NEMS have been recently fabricated and use either an external feedback loop\cite{Kawa} or an intrinsic nanoscale active mechanism\cite{ayari_self_oscillations,Steeneken2011,grogg,Weig2012,1364061,PhysRevB.76.085435,Victor2009}. A self-oscillator (SO) is characterized at the first level by a limit cycle, but its most distinguishing feature is the degree of freedom of its phase which has a zero Lyapunov exponent. The phase can take on any value. Furthermore, contrary to a forced resonator, a forced self-oscillator can keep its phase liberty or enter a synchronized regime depending on the synchronization signal. Synchronization is only possible in systems demonstrating self-oscillations.

Synchronization \cite{pikovsky2003synchronization,balanov2009synchronization} or phase locking, appears in a large variety of systems such as neural networks, lasers, charge density waves, Josephson junction arrays, heart/breathing systems and population of flashing fireflies, and it is expected to be exploited for the treatment of Parkinson's disease, signal processing or opto-mechanical systems \cite{PhysRevLett.109.233906,PhysRevLett.96.103901}. Recently, synchronization has been demonstrated experimentally in NEMS\cite{synchroayari,PhysRevLett.112.014101}. A NEMS due to its small size often operates in a high driving regime. In this regime, synchronization experiments in lasers\cite{PhysRevLett.107.104101}, hydrodynamics\cite{FLM:9059736} and thermoacoustic oscillators\cite{AmJPhys} showed an intriguing phase behavior, never observed before, sometimes called phase trapping\cite{Aronson1990403}. In this article, the full spectrum of behavior of the phase in the high driving regime is unraveled, including phase trapping, thanks to the high phase sensitivity of our self-oscillating nanowires. We define the range were the generation of phase modulated signals is possible and show that the power spectrum density is not the most appropriate tool to define the boundary between each regime.

\begin{figure}
\includegraphics[width=12cm]{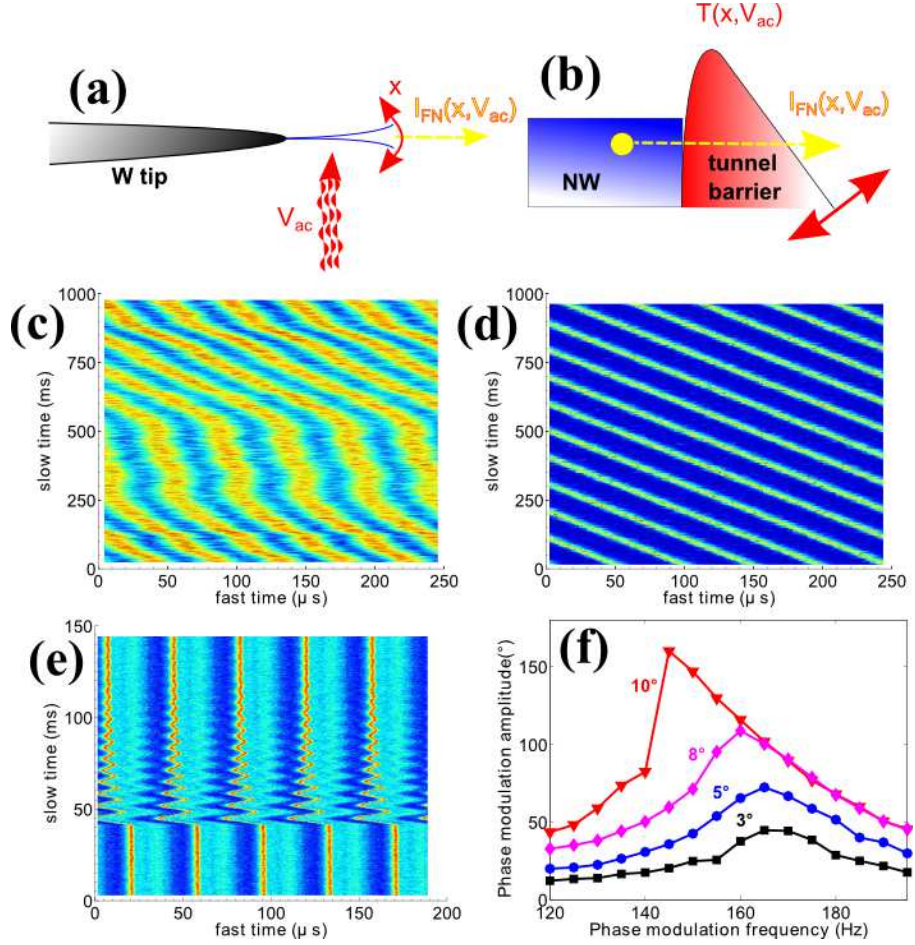}
\caption{\label{fig1} (a) Schematic of a vibrating field emitting nanowire on a tungsten tip submitted to an AC voltage V$_{AC}$. (b) Schematic of the tunneling barrier at the nanowire apex. (c) Temporal map of the field emission current for sample NW1 in the SO regime for V$_{AC}$ = 0 V. The oscillating current is in arbitrary units and the colorscale is such that blue is for low current and red for high current. The averaged current is 600 pA and  the amplitude of current oscillations is in the 100 pA range. (d) Field emission current temporal map for sample NW2 in the synchronized regime for V$_{AC}$ = 5 V and f = 20.8 kHz. The averaged current is 500 pA and  the amplitude of current oscillations is in the 100 pA range.(e) Field emission current temporal map for sample NW1 in the synchronized regime for a driving amplitude V$_{AC}$ = 300 mV,  a driving frequency f$_d$ = 30.9 kHz and a 120$^\circ$ phase jump of the driving after 40 ms. A fit of the relaxation time gives 22.5 ms and an oscillation frequency of the phase of 150.3 Hz. The averaged current is 3.15 nA and  the amplitude of current oscillations is in the 100 pA range. The asymmetry of the current is related to non-linear terms that will be neglected in the rest of the article. (f) Experimental vibration amplitude of the phase of the driven self-oscillator as a function of the phase modulation frequency $\omega_e/2\pi$ for frequency modulation amplitudes ranging from 3$^\circ$ to 10$^\circ$ for NW1.}
\end{figure}

We studied 3 different SiC self-oscillating nanocantilevers (samples NW1, length L = 198 $\mu$m, radius r = 160 nm; NW2, L = 90 $\mu$m, r = 100 nm; NW3, L = 220 $\mu$m, r = 115 nm) fixed at one end to W rigid tips and submitted to AC and DC external electrostatic forces (Fig. \ref{fig1}(a)). The measurements were performed in an ultra high vacuum chamber equipped with a scanning electron microscope column and piezo-inertial nanomanipulators to position the sample in front of a counter electrode at a submicron distance. The motion of the nanowire is transduced into current due to the dependence of the field emission current on the sample to counter electrode distance (for strong amplitude, current rectification by the motion can be visible). The current is collected by a secondary electron detector and recorded on a fast digital oscilloscope. The DC voltage generates a strong electric field at the free end of the nanowire (NW) allowing conduction electrons to tunnel into the vacuum through the field emission triangular barrier. Local electric field variations at the NW apex modulates the transparency of the barrier (Fig. \ref{fig1}(b)). Their origin can be from geometrical changes such as the mechanical displacement of the NW free end as well as from modulated power supplies that induce changes of the voltage on the NW. An intrinsic feedback loop is created because a tiny fluctuation of the position of the NW end changes the transparency of the barrier and the tunneling current, that in turns changes the voltage drop along the NW and the voltage at the apex. The voltage changes modify the transparency of the barrier and the electrostatic force on the NW and thus induce a counter reaction on the NW position. This generates self-oscillation  at a frequency close to the original resonant frequency of the mechanical resonator\cite{PhysRevBself}.

Fig. \ref{fig1}(c), (d) and (e) are measurements of the field emission current amplitude in the self-oscillation regime as a function of two time scales\cite{PhysRevLett.76.2686,PhysRevA.45.R4225}. The maps are obtained as follow : for a fixed V$_{DC}$ the field emission current is recorded for a time between 0.2 s and 1 s depending on the experimental run. The first line of the map is the signal from the beginning to a fixed time $\tau$, where $\tau$ is chosen such that a few oscillation periods are visible on this line. The intensity of the current is represented in colorscale. The second line of the map represents the signal just after $\tau$ up to 2$\tau$ and the rest of the map is build in the same manner. In the absence of phase drift, straight lines of the same color are expected. The angle of these lines depends on the oscillation period and $\tau$.

In Figure \ref{fig1}(c) when no AC signal is applied, on the short time scale (i.e. x axis) the signal appears periodic but on the long time scale (y axis) it  can be seen that the phase has no preferred value and drifts freely like a Brownian  particle.  When an AC voltage is applied with frequency close to the self-oscillating frequency, the self-oscillator is locked and the phase is stabilized (Fig. \ref{fig1}(d)) as can be seen from the very straight lines. Actually, the phase still drifts due to the intrinsic drift of the AC generator, but on a much longer time scale not visible here.

Synchronization results from the competition between the natural frequency of a self-oscillator and the frequency of an external drive. If the external frequency is sufficiently close to the self-oscillation frequency, the self-oscillator is entrained by the external drive. In this synchronized region the self-oscillation frequency disappears. Out of the synchronized region, both frequencies coexist and the system is said to be quasiperiodic\cite{PhysRevA.59.3941,IEEElib}. The dynamics of the phase and amplitude of an individual self-oscillator in the simplest case is governed by a first order time derivative and thus is by nature overdamped\cite{PhysRevBself}. For an abrupt change of the phase of the generator the phase of the self-oscillator should relax exponentially to a new phase value matching the one of the generator like an overdamped particle (OP) relaxing to a potential minimum. Fig. \ref{fig1}(e) shows instead damped oscillations of the phase towards its new fixed value. The phase itself behaves like a resonator. We term this the phase inertia (PI) regime. The open loop relaxation time of the order of 170 ms, the closed loop relaxation time of the order of 800 ms and the detuning frequency of about 1 kHz do not match with the oscillation frequency of the phase of 175 Hz nor the phase relaxation time around several tens of ms. Resonance curve of the phase can be performed if a frequency modulated signal $\varphi_e(t)= \delta\theta\cos(\omega_et)$ is applied to the nanowire where $\delta\theta$ is the frequency modulation amplitude and $\omega_e$ is the swept angular frequency modulation of the phase. Fig. \ref{fig1}(f) is a plot of resonance curves of the phase for different $\delta\theta$. For high forcing $\delta\theta$, the resonance curves shows Duffing non-linearities. For higher forcing the phase unlocks when its amplitude is above 180$^\circ$ and we didn't observe a partly entrained regime. The Duffing non-linearities comes from the $\cos(\varphi)$ term in the phase equation (see supporting information). This behavior is observed for a strong AC driving inducing non-linear oscillations of our samples and a large detuning (i.e. difference between the driving and the NW resonant frequencies). The detuning can typically be up to a few percent of the resonant frequency.

\begin{figure}
\includegraphics[width=12cm]{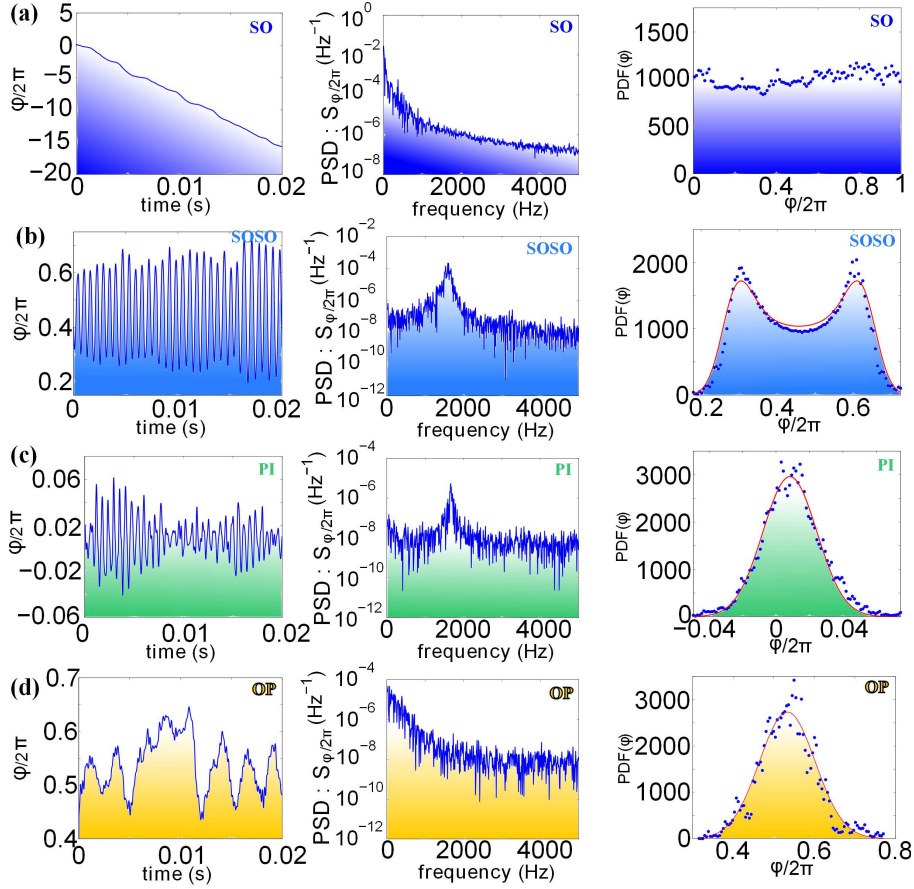}
\caption{\label{fig2} Experimental signature of the phase dynamics in the SO, SOSO, PI and OP regimes for NW 2, (left) time dependence of the phase, (middle) semilog plot of the PSD of the phase and (right) PDF of the phase. (a)  Self-oscillating (SO) regime for V$_{AC}$ = 0.1 V and a detuning of 1 kHz. (b) Self-oscillating self-oscillation regime (SOSO) for V$_{AC}$ = 5.9 V and a detuning of -3 kHz. (c) Phase inertia (PI) regime for V$_{AC}$ = 6.1 V and a detuning of -1 kHz. (d) Overdamped particle regime (OP) for V$_{AC}$ = 0.4 V and a detuning of 0 kHz.}
\end{figure}

This regime of self-oscillation under strong driving and detuning was explored in more detail and four main different dynamical behaviors were observed which are summarized in Fig. \ref{fig2}. Outside the synchronization region, that is for a high enough detuning, the phase difference between the drive and the field emission oscillating current increases quasi linearly (Fig. \ref{fig2} (a)) and the slope is given by the detuning as expected for an unlocked self-oscillator (SO regime). The phase has no preferred value as observed in the probability density function and only  1/f noise appears in the power spectrum density (PSD). For low detuning, the phase is locked and fluctuates around an average value, with fluctuations lower than 2$\pi$ as seen in Fig. \ref{fig2}(d). This is the OP regime mentioned above and the PSD is similar in shape  but smaller in amplitude than for the SO regime. When the detuning is increased, the phase still has gaussian amplitude fluctuations (Fig. \ref{fig2}(c)) around a fixed value like the OP regime but these fluctuations happen with non zero average frequency as observed in the PSD. This is the PI regime. Finally, for a higher detuning, before unlocking, the system can reach a phase modulated regime where the phase enters self-oscillation itself. This regime is called phase trapping\cite{Aronson1990403} or imperfect phase locking\cite{AmJPhys}, in the following we will refer to it as self-oscillating self oscillation (SOSO). The amplitude of phase oscillations in the SOSO regime is roughly constant as observed in Fig. \ref{fig2}(b) and of fixed frequency.

Distinguishing between the SOSO, PI and SO regimes is experimentally challenging because of noise and that is perhaps why the PI has not been observed in the past. The lack of information in the literature makes difficult to compare our noise amplitude with the one of other experiments. Indeed, noise can induce important amplitude fluctuations in the SOSO regime which render the signal for low amplitude of phase self-oscillation similar to the PI regime. When the amplitude of phase self-oscillation becomes large enough to be above the noise level, it may become greater than 2$\pi$ leading to phase unlocking similar to the SO regime. For instance, in \cite{PhysRevLett.107.104101}, the authors observe an enlargement of the phase probability density function (PDF) and interpreted it as the first observation of frequency locking, without phase locking but other interpretations are possible. An increase of the width could also be related to chaotic behavior, increased noise in the PI state, a softening of the restoring force of the phase or phase slips (see supporting information). In Fig. \ref{fig2}(b) a clear signature of SOSO is shown with a typical 2 peak PDF shape indicating that the phase is oscillating with a rather constant amplitude. The PSD itself can not distinguish between the different phase regime.

These four behaviors are generic of a strongly driven self-oscillator and can theoretically be obtained even from a simple driven Van der Pol oscillator\cite{pikovsky2003synchronization,balanov2009synchronization}(see supporting information) :
\begin{equation}
\label{Eq1} m\ddot{x}+m\lambda\dot{x}+m\gamma\dot{x}x^2+m\omega_0^2x = F cos(\omega_dt)
\end{equation}
where m is the effective mass of the nanocantilever, x the displacement of the nanocantilever tip in the transverse direction, the dot refers to the derivative versus time t, $\lambda$ is the linear negative damping, $\gamma$ is the non linear damping coefficient responsible for the limit cycle, $\omega_0/2\pi$ is the natural frequency of the resonator, $F$ the electrostatic force and $\omega_d/2\pi$ the frequency of the external driving. The driven self-oscillator position is given by $x(t) = R \cos(\omega_d t -\varphi)$ where $R$ is the amplitude of self-oscillation, and $\varphi$ its phase. $R(t)$ and $\varphi(t)$ are the two slowly varying degrees of freedom compared to the period of the self-oscillator. From Eq.\ref{Eq1}, the amplitude and phase dynamical equations can be deduced :
\begin{equation}
\label{EqTph2} R_0\dot{\widetilde{\varphi}} = \delta \omega \widetilde{R} + \frac{1}{2}(-\lambda-\frac{\gamma}{4}R_0^2)R_0\widetilde{\varphi}
\end{equation}
\begin{equation}
\label{EqTR2} \dot{\widetilde{R}} = \frac{1}{2}(-\lambda-\frac{3\gamma}{4}R_0^2)\widetilde{R}-\delta \omega R_0\widetilde{\varphi}
\end{equation}
where $\delta \omega = (\omega_d^2-\omega_0^2)/2\omega_d$ can be assimilated to the detuning frequency. $R_0$ is the self-oscillation amplitude, $\widetilde{\varphi}$ a small perturbation around the equilibrium phase and $\widetilde{R}$ a small perturbation around $R_0$ (The equilibrium phase and amplitude value can be obtained from equations given in supporting information). After some algebra the phase and amplitude dynamical equations give this linear equation for the phase :
\begin{equation}
\label{Eqphext1bis} \ddot{\widetilde{\varphi}} + \Gamma\dot{\widetilde{\varphi}}+\Omega^2\varphi = 0
\end{equation}
where $\Gamma = \lambda+\frac{\gamma}{2}R_0^2$ and $\Omega^2 = \frac{1}{4}[(\lambda+\frac{3\gamma}{4}R_0^2)(\lambda+\frac{\gamma}{4}R_0^2)+ 4\delta \omega^2$].
From this equation, we get the OP regime for low detuning and the PI regime for higher detuning. For even higher detuning $R_0$ can decrease enough so that the damping $\Gamma$ becomes negative thus reaching the SOSO regime. Non-linear terms can be added to this equation to model the SOSO to SO transition\cite{balanov2009synchronization} (see supporting information). This model predicts that for high detuning, the phase oscillation frequency $\Omega$ is independent of the forcing amplitude and is equal to the detuning.

\begin{figure}
\includegraphics[width=12cm]{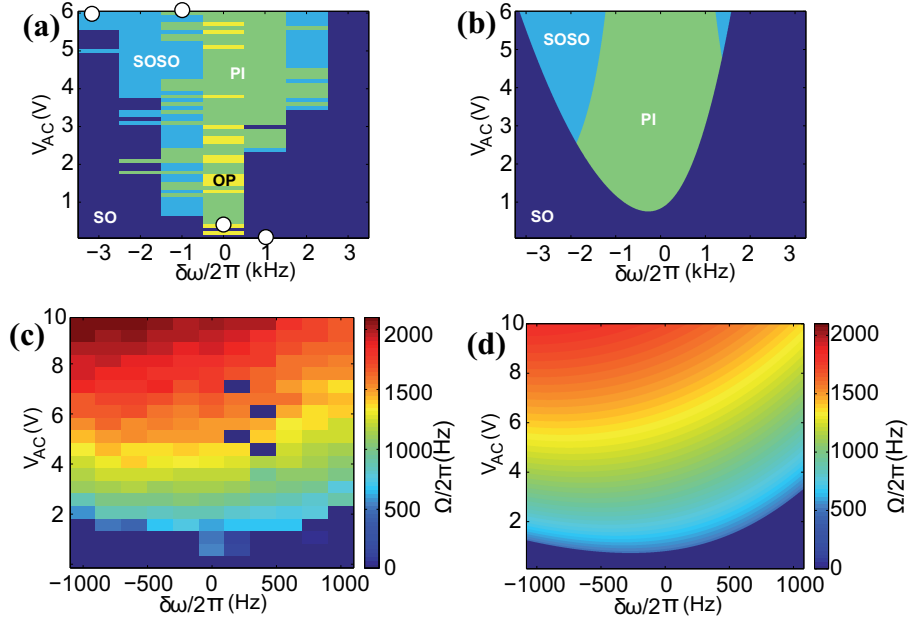}
\caption{\label{fig3} (a) Experimental frequency-amplitude mapping (i.e Arnold tongue) of the different synchronization regime for NW2. The white dots corresponds to the experimental points of Fig. \ref{fig2}. Dark blue SO regime, light blue SOSO regime, green PI regime and yellow OP regime. (b) Simulated frequency-amplitude mapping of the different synchronization regime. (c) Experimental frequency-amplitude mapping of the phase frequency $\Omega/2\pi$ in the PI region for the same experimental conditions as in (a). (d) Simulated frequency-amplitude mapping of the phase frequency $\Omega/2\pi$ in the PI region.}
\end{figure}

We performed a detuning-amplitude mapping of the synchronization region. In Fig. \ref{fig3}(a), a point in the map is considered as in the SO regime if the phase is not bounded, in the SOSO regime if the PDF of the phase has 2 peaks, in the PI regime if the PSD has a peak for a frequency different than zero as well as the PDF of the phase has a single peak and in the OP regime otherwise. In Fig. \ref{fig3}(c) the phase frequencies are extracted from the maximum of the phase PSD. Due to fluctuations, the exact frontier of the so-called Arnold tongue\cite{pikovsky2003synchronization,balanov2009synchronization} may vary but the 4 regimes are always present when the measurement is repeated and the SOSO regime always appears for high forcing and detuning. The points marked as OP may be either true OP states or more likely PI states with the amplitude of the 1/f noise dominating the oscillation frequency. Besides the existence of a large SOSO region, the most remarkable feature of this tongue is the behavior of the phase frequency (Fig. \ref{fig3}(c)). This frequency is strongly dependent on the forcing amplitude and is different from the detuning frequency contrary to the Van der Pol model predictions above. Thus it is not directly related to frequency beating between the two main signals. We have developed a phenomenological model to explain this behavior by adding non-linear dependencies of the damping and the frequency in the Van der Pol model. $\omega_0$ is replaced in Eq. \ref{Eq1} by : $\omega_t^2 = \omega_0^2 + \mu^*(\omega_d^2-\omega_0^2)x^2 + \nu\sqrt{F}$ where $\nu$ is a negative constant that takes into account the fact that the maximum of the phase frequency shift towards negative detuning for increasing F and $\mu^* = \mu/R_0^2$ is such that this term is less effective for high self-oscillation amplitude $R_0$. The physical origin of such non-linearities could be the strong electrostatic frequency tuning of nanocantilevers during field emission \cite{PhysRevLett.89.276103} and the non-linear intrinsic feedback between the electrical circuit and mechanical motion \cite{lazarus:193114,ayari_self_oscillations}. The non-linearities of the field emission current down-mixes the driving voltage and the self-oscillation signal (the term $\omega_d^2-\omega_0^2 \approx 2(\omega_d-\omega_0)\omega_0$ is proportional to the detuning). This mixed signal is then transmitted to the apex voltage due to the voltage drop across the nanowire which in turns modifies the tuning frequency. If the electrical circuit has an in phase back action on the mechanical motion, it will have also an out of phase effect and so the damping will be affected too. This induces that the damping should have an expression of the form $\lambda = \lambda_0 +\lambda_1F(\omega_d-\omega_0)^2$ where $\lambda_0$ and $\lambda_1$ are constant. This model gives a very good agreement with experimental data (Fig. \ref{fig3} (b)-(d)).

In conclusion, we have observed experimentally low and even negative friction phase motion in a field emission nano-electro-mechanical system. We found that the phase itself has a resonant frequency when driven at high amplitude and can self-oscillate for high mismatch between the self-oscillator frequency and the driving frequency. In principle, synchronization of the self-oscillating phase and even observation of a "self-oscillation of the phase of the self-oscillating phase" should be possible. In our experiment, this regime was too unstable as this would require additional phase motion in a region close to the desynchronization limit. However, such a phenomenon may be observable by working at low temperature thus considerably reducing the phase noise. Although NEMS and this system in particular are not mature enough for radio communication applications, the regime of SOSO could be an original way of generating frequency modulated signals for better data transmission \cite{ayariFM} and signal processing. This work was limited to the study of a single self-oscillator in the high driving limit but SOSO, which was challenging to observe previously in isolated oscillators, might be an interesting phenomenon to study in the dynamics of multiple coupled self-oscillators in the strong coupling limit. The Adler equation \cite{Adler} and the Kuramoto model \cite{kuramoto2003chemical} are widely used to describe the dynamics of synchronized systems but take into account only the phase degree of freedom. Phase resonance and phase self-oscillation can be theoretically introduced into the Kuramoto framework by inserting phenomenological inertial terms and phase delay between several self-oscillators \cite{PhysRevE.65.026208}. However strong coupling is assumed which is incompatible with the Kuramoto model\cite{RevModPhys.77.137}. In strongly non-linear self-oscillating systems such as NEMS, the phase and amplitude are coupled and the Adler equation or the Kuramoto model are no longer valid. In particular, multiple self-oscillating field emission NEMS coupled electrostatically could be an interesting system for studying the transition from locking to the unsteady regime of synchronization with Hopf oscillation and then amplitude death as predicted in \cite{PhysRevLett.65.1701} for coupled self-oscillators.

The authors acknowledge the "Plateforme Nanofils et Nanotubes Lyonnaise" of the University
Lyon1. This work was supported by the R\'egion Rh\^one-Alpes (PROGRAMME SRESR - CIBLE 2008 and cluster MACODEV) and the French National Research
Agency (ANR) through its Nanoscience and Nanotechnology
Program (NEXTNEMS, ANR-07-NANO-008-01 and NEMSPiezo, ANR-08-P078-48-03) and Jeunes
Chercheuses et Jeunes Chercheurs Program (AUTONOME,
ANR-07-JCJC- 0145-01).

\bibliography{synchnl}

\end{document}